\documentclass[final, authoryear, 5p, times, twocolumn]{elsarticle}
\usepackage{graphicx}
\journal{New Astronomy}

\def\astrobj#1{#1}

\begin{document}
\begin{frontmatter}

\title{Physical Parameters of Some Close Binaries: ET Boo, V1123 Tau, V1191 Cyg, V1073 Cyg and V357 Peg}

\author[AU]{F. Ekmek\c{c}i\corref{cor1}}
\author[AU]{A. Elmasl{\i}}
\author[AU]{M. Y{\i}lmaz}
\author[AU]{T. K{\i}l{\i}\c{c}o\u{g}lu}
\author[NU]{T. Tanr{\i}verdi}
\author[AUO]{\"{O}. Ba\c{s}t\"{u}rk}
\author[AU]{H. V. \c{S}enavc{\i}}
\author[AU]{\c{S}. \c{C}al{\i}\c{s}kan}
\author[AU]{B. Albayrak}
\author[AU]{S. O. Selam}

\cortext[cor1]{e-mail: fekmekci@science.ankara.edu.tr}
 
\address[AU]{Ankara University, Faculty of Science, Department of Astronomy and Space Sciences, TR-06100, Tando\u{g}an, Ankara, Turkey}
\address[NU]{Ni\u{g}de University, Faculty of Arts and Sciences, Department of Physics, 51240, Ni\u{g}de, Turkey}
\address[AUO]{Ankara University Observatory, 06837, Ahlatl{\i}bel, Ankara, Turkey}

\begin{abstract}

With the aim of providing new and up-to-date absolute parameters of some close binary systems, new {\it BVR} CCD photometry was carried out at the {\it Ankara University Observatory (AUG)} for five eclipsing binaries, \astrobj{ET Boo}, \astrobj{V1123 Tau}, \astrobj{V1191 Cyg}, \astrobj{V1073 Cyg} and \astrobj{V357 Peg} between April, 2007 and October, 2008. In this paper, we present the orbital solutions for these systems obtained by simultaneous light and radial velocity curve analyses. Extensive orbital solution and absolute parameters for \astrobj{ET Boo} system were given for the first time through this study. According to the analyses, \astrobj{ET Boo} is a detached binary while the parameters of four remaining systems are consistent with the nature of contact binaries. The evolutionary status of the components of these systems are also discussed by referring to their absolute parameters found in this study.

\end{abstract}


\end{frontmatter}


\section{Introduction}

Eclipsing binary stars are essential key objects in the field of stellar astrophysics to reach precise absolute stellar dimensions (i.e., masses, radii etc.) and laboratories to test different theoretical aspects on stellar structure and evolution.

Obtaining absolute parameters of close binary systems is very important in examining/supporting  the general consideration of the evolutionary history of close binary stars \citep[see e.g.,][]{kraic87,kraic88,zahn89,zahnb89,fran92}. 
In this context, we present the results of Wilson-Devinney (WD) light curve analysis of \astrobj{ET Boo}, \astrobj{V1123 Tau}, \astrobj{V1191 Cyg}, \astrobj{V1073 Cyg} and \astrobj{V357 Peg} based on new {\it BVR} CCD observations carried out at the {\it Ankara University Observatory (AUG)} and spectroscopic observations by  Rucinski's research team published in a series of fifteen papers on "Radial Velocity Studies of Close Binary Stars". The current information on these five systems could be summarized as follows:

\astrobj{ET Boo} (SAO 45318, BD +47$^{\circ}$  2190, HIP 73346, 2MASS $J14592031+4649036$)  is a  $\beta$ Lyr type eclipsing binary, which is also a member of a quadruple system \citep{pribul06}, and found to be a close visual binary system in 1978 by \citet{couet81}. The spectral type of the system was given as F8 by \citet{bart02}. Up to now, the photometric observations of the system have appeared to be rather unattended with the exception of the study by \citet{oja86} which only gives colour indices of ${\it B-V}= 0.60$ and ${\it U-B}= - 0.01$ for \astrobj{ET Boo}.

\astrobj{V1123 Tau} (BD +17$^{\circ}$  579, HIP 16706, ADS2624A) was discovered during the HIPPARCOS mission \citep{esa97} as a $\beta$ Lyr type eclipsing binary, and it was classified as W UMa type eclipsing binary by \citet{kaz09}. The first ground-based photometry was made by \citet{ozdar06}. They only gave the light and colour curves of \astrobj{V1123 Tau}, obtained in the years 2003 and 2005, together with the new light elements. Recent radial velocity studies for a number of close binary stars, including \astrobj{V1123 Tau}, were carried out by \citet{rucin08}. They pointed out that \astrobj{V1123 Tau} is accompanied by a fainter companion (ADS2624B, $\rho = 4.3^{"}$, $\theta = 136^{\circ}$, and $\Delta V = 1.77$), and the light contribution of this visual companion appears to have affected the light amplitudes of the system.  \citet{guti09} presented the results of optical spectroscopy of the visual binary \astrobj{ADS2624} and confirmed that the spectral types of components A (\astrobj{V1123 Tau}) and B are G0V and K0V, respectively. The system has been classified as a W-subtype of W UMa systems by \citet{rucin08}. \citet{deb11} performed a light curve analysis for V band observations of ASAS-3 project using Wilson-Devinney code. Separately,
\citet{zahng11} presented the absolute parameters of \astrobj{V1123 Tau} system based on their Wilson-Devinney analysis for the light curves obtained with the high precision, multi-band CCD observations of this binary system. 

\astrobj{V1191 Cyg} (GSC 03159-01512, 2MASS J20165081+4157413, TYC 3159-1512-1) was found to be a variable star by \citet{mayer65}. \citet{pribul05a} gave new CCD observations in {\it BVRI} filters, and carried out a light curve analysis using Pribulla's ROCHE code. They performed a grid search in the mass ratio (q) parameter during the analysis since no spectroscopic mass ratio was known for the system. They defined the system as W-subtype and gave the results for geometric elements as ${\it i} = 80.4(4)^{\circ}$, $q = 0.094$, $f = 0.46(2)$. Based on the systematic deviation of the fit in R passband around primary minimum they pointed out a probability of the presence of a third companion. Recently, \citet{ulas12} presented the parameters for the hot and cooler companions together with the photometric and spectroscopic variations of \astrobj{V1191 Cyg} binary system. This system was also studied by \citet{zhu11} using their CCD photometric light curves in {\it BV(RI)c} bands obtained in 2009. They derived the absolute parametrs of \astrobj{V1191 Cyg} based on their spectroscopic and photometric solutions.

\astrobj{V1073 Cyg} (BD +33$^{\circ}$  4252, HD 204038, HIP 105739) is a W UMa type eclipsing binary system. The variability of \astrobj{V1073 Cyg} was first recognized by \citet{stroh60}. \citet{stroh62} published photographic minima of the system together with the first light elements and a photographic light curve. \citet{fitzg64} obtained first radial velocity curve and found the mass ratio (q) to be 0.34 and spectral classification A3~Vm for the primary component. \citet{kondo66} solved the light curves of the system in {\it Y} (5410 $\AA$) and {\it B} (4250 $\AA$) bands using Russell-Merrill technique and proposed a contact model for the system. \citet{krus67} obtained light curves in blue and yellow bands for which they haven't given the effective wavelengths. They also well fitted the data given by \citet{kondo66} based on their results. \citet{abtb69} classified the system as F0n~III-IV or F0n~V, contradicting A3m~V classification of \citet{fitzg64} that used only hydrogen lines. \citet{leusch78} analyzed the light curves obtained by \citet{kondo66} with the Wilson-Devinney computer code. They have proposed that \astrobj{V1073 Cyg} was an overcontact system with a fill-out factor of 7\%. \citet{niar78} reanalyzed the light curves of both \citet{kondo66} and \citet{bend67} making use of Kopal's method. \citet{aslher84} analyzed the orbital period behavior of the system and detected a sudden period decrease of 0.4 seconds in 1976. Sezer \citep[see][]{sezer93,sezer94}, found 8\% overcontact and a mass ratio of 0.436 assuming $T{_1}=8570 K$. \citet{wolfd92} gave an O-C diagram and found that the period was constantly decreasing. \citet{ahn92} analyzed three sets of previously published light curves with the program LIGHT2, assuming convective behavior, and $T_{1}=6700 K$ (consistent with the previously derived spectral type of F2), and found 8\% overcontact with both components near the terminal-age main sequence (TAMS). \citet{mornaft00} observed \astrobj{V1073 Cyg} for 4 consecutive nights in July 1998 and analyzed the light curves they obtained, adopting the mass ratio published by \citet{fitzg64} and using parameters found by \citet{sezer96} as initial parameters. \citet{mornaft00} indicated that most of the photometric analyses until the time of their study relied on BD +33$^{\circ}$ 4248 as comparison star which they tested its variability and found no evidence. \citet{mornaft00} made an orbital period analysis as well, and found that mass transfer was a more plausible explanation for the orbital period change than mass loss, because of the system's high escape velocity. Another period analysis of the system was published by \citet{yanli00} who also reported that the orbital period was not stable but decreasing and that the light curves of the system show unstable behavior too.  They detected a positive O'Connell effect in \citet{kondo66}'s light curves while a negative O'Connell effect was observed by \citet{sezer93}. \citet{pribruc06} and \citet{rucin07} could not detect an additional component using adaptive optics observations. The system is also defined as an A-subtype of W UMa systems \citep{pribul06}.

\astrobj{V357 Peg} (BD +24$^{\circ}$ 4828, HD 222994, HIP 117185, SAO 91468) was discovered and classified as W UMa type binary system during the HIPPARCOS mission \citep{esa97}. The first photometric light curves of the system were obtained by \citet{yasar00} but no analyses were performed. \citet{selam04} analysed the HIPPARCOS light curve of \astrobj{V357 Peg} with Rucinski's simplified light curve synthesis method \citep{rucin93} and derived the mass ratio and inclination of the system as 0.30 and 75$^{\circ}$, respectively. The first radial velocity curve was given by \citet{rucin08}. They performed the first spectroscopic observation and concluded that \astrobj{V357 Peg} is an A-subtype contact binary system with a mass ratio of 0.401 and spectral type of F2~V by using the spectra taken between 1997 and 2005. Recently, \citet{deb11} presented the results for the \astrobj{V357 Peg} binary system by using V band observations of ASAS-3 project in their Wilson-Devinney light curve analysis. 

\section{Observations}

CCD observations of five eclipsing binaries (\astrobj{ET Boo}, \astrobj{V1123 Tau}, \astrobj{V1191 Cyg}, \astrobj{V1073 Cyg} and \astrobj{V357 Peg}) were carried out by using an Apogee ALTA U47+CCD camera (1024x1024 pixels) with {\it BVR} filters mounted on a 40 cm Schmidt-Cassegrain telescope of the {\it Ankara University Observatory (AUG)}. The log of observations is given in Table~\ref{table1}. The reduction of the CCD frames has been performed with standard packages of IRAF\footnote{IRAF is distributed by the National Optical Astronomy Observatory, which are operated by the Association of Universities for Research in Astronomy, Inc., under cooperative agreement with the National Science Foundation.}, and the individual differential {\it BVR} observations were computed in the sense of variable minus comparison star. The corresponding orbital phases for each variable were computed with the light elements listed in Table~\ref{table2}. The light elements of \astrobj{V1123 Tau}, listed in Table~\ref{table2}, were calculated by using the eight minima given by \citet{yilma09} and three minima from the observations of this study. The light elements of \astrobj{V1191 Cyg} were calculated by using the 11 published minima by various authors \citep{pribul05b,nels06,hubs06,nels07,pari07,hubsibvs07}. The light elements of \astrobj{V1073 Cyg} were calculated by using 19 minima obtained in various studies \citep{muyes96,nels98,nels03,drm03,brat07,brat08,yilma09} and those published in WEB sites (http://astro.sci.muni.cz/variables/ocgate/; www.antonpaschke.com). For \astrobj{V357 Peg} light elements were calculated by using 15 published minima by \citet{kesk00}, \citet{alis02}, \citet{tanri03}, \citet{alb05}, \citet{drozdz05}, \citet{dvor05}, \citet{pari07} and \citet{hubs07}, and 5 minima obtained from the observations of this study. Information for the comparison and check stars, used during the observations, are given in Table~\ref{table3}.

\begin{table}[ht]
\scriptsize
\begin{center}
\caption{The log of CCD observations.}
\label{table1}
\medskip
\resizebox{9cm}{!} {\begin{tabular}{lccccccc}
\hline
System & Obs. dates & \multicolumn{3}{c}{\# of data} & \multicolumn{3}{c}{Nightly mean errors} \\ 
  &  & $\Delta B$ & $\Delta V$ & $\Delta R$ & $\sigma_{B}$(mag) & $\sigma_{V}$(mag) & $\sigma_{R}$(mag) \\
\hline
         ET Boo   & 16 Apr; 4,5,11 May & 562 & 682 & 619 & 0.004 & 0.004 & 0.005 \\
                       &  2007              &    &    &    &    &   &  \\
        V1123 Tau & 9,12,16,30 Sept;   & 644 & 645 & 643 & 0.004 & 0.003 & 0.004 \\
                          & 1 Oct 2008         &   &   &   &   &   &   \\
        V1191 Cyg & 5,6,8,10,12,14,17  & 931 & 932 & 953 & 0.005 & 0.003 & 0.001 \\
                           &  Jul 2008          &    &   &   &   &   &   \\
        V1073 Cyg  & 9,13,18,19 Aug;2,14& 1444  &  1326  &  1324  &  0.003 &  0.007  & 0.008   \\
                           & Sept 2008          &    &    &    &   &    &    \\
        V357 Peg   & 10,16,23 Aug;       &    &    &   &    &    &    \\
                          & 4,13,16,18 Sept;    & 1583 & 1730 & 1867 & 0.001 & 0.001 & 0.001 \\
                          & 8,19 Oct 2008       &   &   &   &   &   &   \\ 
\hline                       
\end{tabular}}
\end{center}
\end{table}

\begin{table}
\scriptsize
\begin{center}
\caption{The light elements used in this study for each system.}
\label{table2}
\medskip
\begin{tabular}{lccc}
\hline
                   & Epoch &       &       \\ 
System       & (HJD+2400000) & Period (days) & Ref. \\
\hline
ET Boo & 52701.5928 & 0.6450398 & Pribulla et al. (2006) \\
V1123 Tau & 54719.5749(5) & 0.3999441(98) & This study \\ 
V1191 Cyg & 54672.3431(19)& 0.3133885(5)  &  This study  \\
V1073 Cyg & 54698.5000(51) & 0.7858500(70) & This study  \\
V357 Peg & 54702.4652(5) & 0.5784511(1) & This study \\ 
\hline
\end{tabular}
\end{center}
\end{table}

\begin{table}[ht]
\scriptsize
\begin{center}
\caption{The comparison and check stars used during the observation for each variables.}
\label{table3}
\bigskip
\begin{tabular}{llll}
\hline
        & Variable  & Comparison & Check  \\
        \hline
                          &  ET Boo                & GSC 03474-00142                  & - \\
        $\alpha_{2000}$: & 14$^{h}$59$^{m}$20$^{s}$.32       & 14$^{h}$59$^{m}$10$^{s}$.00      & - \\
        $\delta_{2000}$: & +46$^{\circ}$49$^{'}$03$^{"}$.61 & +46$^{\circ}$41$^{'}$29$^{"}$.20 & - \\
        Spec. Typ.:       & F8                                & -                                & - \\
        V(mag):           & 9.09                              & 10.9                             & -\\
        \hline
                          & V1123 Tau              & GSC 1238-1039                    & GSC 1238-1028 \\
        $\alpha_{2000}$: & 03$^{h}$34$^{m}$58$^{s}$.55       & 03$^{h}$34$^{m}$28$^{s}$.87      & 03$^{h}$34$^{m}$27$^{s}$.48 \\
        $\delta_{2000}$: & +17$^{\circ}$42$^{'}$38$^{"}$.04 & +17$^{\circ}$42$^{'}$12$^{"}$.18 & +17$^{\circ}$35$^{'}$46$^{"}$.50 \\
        Spec. Typ.:       & G0                               & A5                                & - \\
        V(mag):           & 9.97                             & 10.9                              & 11.44\\
        \hline
                          & V1191 Cyg             & HD 228669                         & HD 228695 \\
        $\alpha_{2000}$: & 20$^{h}$16$^{m}$50$^{s}$.81      & 20$^{h}$16$^{m}$01$^{s}$.92       & 20$^{h}$16$^{m}$18$^{s}$.55 \\
        $\delta_{2000}$: & +41$^{\circ}$57$^{'}$41$^{"}$.36 & +41$^{\circ}$58$^{'}$00$^{"}$.18 & +41$^{\circ}$58$^{'}$53$^{"}$.11 \\
        Spec. Typ.:       & -                                & F8                                & A0 \\
        V(mag):           & 10.8                             & 10.59                             & 9.99\\
        \hline
                          & V1073 Cyg             & GSC 2711-2412                     & GSC 2711-2014 \\
        $\alpha_{2000}$: & 21$^{h}$25$^{m}$00$^{s}$.36      & 21$^{h}$24$^{m}$36$^{s}$.99       & 21$^{h}$24$^{m}$53$^{s}$.80 \\
        $\delta_{2000}$: & +33$^{\circ}$41$^{'}$14$^{"}$.94 & +33$^{\circ}$48$^{'}$21$^{"}$.15  & +33$^{\circ}$50$^{'}$12$^{"}$.59 \\
        Spec. Typ.:       & F1V                            & A0                                & - \\
        V(mag):           & 8.38                             & 8.87                             & 10.81 \\
        \hline
                          & V357 Peg              & GSC 2254-1156                     & GSC 2254-2715 \\
        $\alpha_{2000}$: & 23$^{h}$45$^{m}$35$^{s}$.06      & 23$^{h}$45$^{m}$09$^{s}$.02       & 23$^{h}$45$^{m}$29$^{s}$.94 \\
        $\delta_{2000}$: & +25$^{\circ}$28$^{'}$18$^{"}$.94 & +25$^{\circ}$21$^{'}$12$^{"}$.05 & +25$^{\circ}$21$^{'}$46$^{"}$.20 \\
        Spec. Typ.:       & F5                               & -                                 & - \\
        V(mag):           & 9.06                             & 10.54                             & 11.34\\
\hline
\end{tabular}
\end{center}
\end{table}

\section{Analyses of the light curves}

During the light curve analyses, the PHOEBE graphical user interface developed by \citet{prsa05} to visualize the well known Wilson-Devinney (WD) light curve analysis code \citep{wils71} was used. The adjustable parameters are the inclination $i$, the non-dimensional potentials ($\Omega_{h,c}$), the surface temperature of the components ($T_{ph}$) and the relative monochromatic luminosities ($L$) in each passband. The limb-darkening coefficients of logarithmic law (the values of x and y in Table~\ref{table4}) were taken from \citet{vanha93}'s tables. In the solutions, the rotation axis of the components were taken to be perpendicular to the orbital plane, and synchronized rotation was assumed for the component stars. The initial estimations of the third light contribution for \astrobj{ET Boo} and \astrobj{V1123 Tau} were made by using the corresponding $M_{v}$ taken from Table II of \citet{strai81} on the adopted calibration of MK spectral types in absolute magnitudes $M_{v}$. The differential correction program was initiated for the simultaneous solution of the light and radial velocity curves and then a visual inspection of the agreement between the synthetic and observational light curves was made. The radial velocity (RV) data of all systems were taken from the series of papers on "Radial Velocity Studies of Close Binary Stars"  by Rucinski and his collaborators. RV data of \astrobj{ET Boo} and \astrobj{V1073 Cyg} were taken from \citet{pribul06}, and RV data of \astrobj{V1123 Tau}, \astrobj{V1191 Cyg}, and of \astrobj{V357 Peg} were taken from \citet{rucin08}. The goodness of fits $\Sigma (O-C)^{2}$ to the light curves were checked for every run. After reaching a satisfactory agreement, we fixed the parameters mentioned above and adjusted for the various spot parameters if necessary. The results are summarized in Table~\ref{table4}. The temperatures are given in units of Kelvin degrees and the longitudes ($\lambda$), latitudes ($\beta$) and the radii (r) of spots are in arc degrees. The "latitude" of a star spot center, measured from 0 degrees at the "north" (+ z) pole to 180 degrees at the "south" pole of the star. And, the longitude of a star spot center, measured counter-clockwise (as viewed from above the + z axis) from the line of star centers from 0 to 360 degrees \citep{wilso93}. From Table~\ref{table4}, it can be seen that the achievement of the results of spot parameters of the system were obtained with no spots for \astrobj{ET Boo}, one dark spot located on the cooler component of \astrobj{V1123 Tau}, two dark spots located on the cooler component of \astrobj{V1191 Cyg}, one dark spot located on the cooler component of \astrobj{V1073 Cyg},  and one bright spot located on the hotter component of \astrobj{V357 Peg}. \citet{rucin08} had determined the large photospheric dark spot to be probably on the seconadry component of \astrobj{V357 Peg} around 0.75 orbital phase by using their observations made between August 25 and September 6, 2005. They also pointed out that the observatios made in 1997 did not show any indication of the photospheric spot. Therefore, at first the WD analysis of the light curves of \astrobj{V357 Peg} was attempted to be run with the dark spot on the secondary component of the system, but no suitable result was achieved. Then, it was seen that the analysis with bright spot on the hotter component gave a satisfactory result for \astrobj{V357 Peg}.


\begin{table}
\scriptsize
\begin{center}
\caption{Results of simultaneous WD code analysis of {\it BVR} CCD observations of five close binary systems.}
\label{table4}
\medskip
\resizebox{9cm}{!} {\begin{tabular}{lccccc}
\hline
System & ET Boo & V1123 Tau & V1191 Cyg & V1073 Cyg & V357 Peg \\
\hline
q($M_{c}$/$M_{h})$      & 0.884	& 3.584	& 9.346 	& 0.303 	& 0.401 \\
e (eccentricity)        & 0             & 0             & 0             & 0             & 0      \\
$V_{\gamma}$(kms$^{-1}$)& - 23.35       & 25.32         & - 16.82       & - 6.85        & - 11.026 \\
i($^{\circ}$)	& 76.3$\pm$0.1 & 74.01$\pm$1.05 & 83.2$\pm$2.2 & 69.85$\pm$0.72 & 73.34$\pm$1.57 \\
\hline
Component		& hot/cool	&hot/cool	&hot/cool	&hot/cool	&hot/cool   \\
\hline
$\Omega$		&4.657$\pm$0.006&7.194$\pm$0.005&14.144$\pm$0.05&2.44$\pm$0.08&2.60$\pm$1.44 \\
                        &3.817$\pm$0.004&7.194$\pm$0.005&14.144$\pm$0.05 &2.44$\pm$0.08&2.60$\pm$1.44 \\
$T_{ph} (K)$		&6125   	&5920	        &6300 	        &6700 	        &7000  \\
                        &5758$\pm$40    &5821$\pm$31    &6215$\pm$55    &6520$\pm$74    &6687$\pm$971 \\
Albedo			&0.5/0.5	&0.5/0.5	&0.5/0.5	&0.5/0.5	&0.65/0.65 \\
g (gravity dark.)	&0.32/0.32	&0.32/0.32	&0.32/0.32	&0.32/0.32	&0.35/0.35 \\
x({\it B})	        &0.822/0.837	&0.830/0.834	&0.349/0.371	&0.798/0.804	&0.188/0.257 \\
x({\it V})	        &0.736/0.760	&0.749/0.756	&0.139/0.154	&0.702/0.710	&0.062/0.095 \\
x({\it R})	        &0.643/0.668	&0.657/0.664	&0.027/0.042	&0.608/0.616	&-0.043/-0.016 \\
y({\it B})	        &0.197/0.141	&0.166/0.151	&0.536/0.516	&0.255/0.240	&0.694/0.624 \\
y({\it V})	        &0.261/0.233	&0.248/0.239	&0.674/0.664	&0.283/0.278	&0.724/0.701 \\
y({\it R})	        &0.271/0.253	&0.263/0.257	&0.695/0.687	&0.287/0.284	&0.734/0.719 \\
$r_{pole}$		&0.263/0.315	&0.268/0.473	&0.201/0.536	&0.463/0.270	&0.447/0.298 \\
$r_{point}$		&0.277/0.365	&---/---	&---/---	&---/---	&---/--- \\
$r_{side}$		&0.267/0.326	&0.281/0.512	&0.210/0.599	&0.499/0.282	&0.481/0.313 \\
$r_{back}$		&0.273/0.346	&0.324/0.541	&0.249/0.619	&0.527/0.321	&0.513/0.358 \\
\hline
$\lambda_{spot \#1}$($^{\circ}$)&---/--- &---/237.56	&---/276.17     &---/333.14	&310/---\\
$\lambda_{spot \#2}$($^{\circ}$)&---/--- &---/---	&---/166.35	&---/---	&60/---	\\
$\beta_{spot \#1}$($^{\circ}$)	&---/--- &---/91.57     &---/102.29	&---/15.09		&50/---	\\
$\beta_{spot \#2}$($^{\circ}$)	&---/--- &---/---       &---/90.15	&---/---	&60/---	\\
r$_{spot \#1}$($^{\circ}$)      &---/--- &---/11.69     &---/8.16	&---/50.66	        &28/---	\\
r$_{spot \#2}$($^{\circ}$)      &---/--- &---/---       &---/9.89	&---/---	&---/--- \\
f(T)$_{spot \#1}$	        &---/---&---/0.784      &---/0.786	&---/0.879	&1.085/--- \\
f(T)$_{spot \#2}$	        &---/---&---/---        &---/0.838	&---/---	&---/--- \\
\hline
f (fillout)             &   ---        & 0.165     &  0.295     &   0.174      &   0.312 \\
$L_{1}$/($L_{Total}$)(in {\it B}) &0.366 &0.220   &0.133  &0.933 &0.741 \\
$L_{1}$/($L_{Total}$)(in {\it V}) &0.345 &0.211   &0.130  &0.928 &0.730 \\
$L_{1}$/($L_{Total}$)(in {\it R}) &0.329 &0.210   &0.128  &0.921 &0.723 \\
$L_{2}$/($L_{Total}$)(in {\it B}) &0.389 &0.609   &0.868  &0.077 &0.258 \\
$L_{2}$/($L_{Total}$)(in {\it V}) &0.391 &0.602   &0.870  &0.072 &0.269 \\
$L_{2}$/($L_{Total}$)(in {\it R}) &0.392 &0.610   &0.872  &0.079 &0.277 \\
$L_{3}$/($L_{Total}$)(in {\it B})   &0.245$\pm$0.002 &0.171$\pm$0.002 & --- & --- & --- \\
$L_{3}$/($L_{Total}$)(in {\it V})   &0.264$\pm$0.002 &0.186$\pm$0.002 & --- & --- & --- \\
$L_{3}$/($L_{Total}$)(in {\it R})   &0.278$\pm$0.002 &0.179$\pm$0.002 & --- & --- & --- \\
\hline
$\Sigma (O-C)^{2}$ (in {\it B})   &0.019 &0.053 &0.077 &0.069 &0.259 \\
$\Sigma (O-C)^{2}$ (in {\it V})   &0.045 &0.033 &0.043 &0.141 &0.106 \\
$\Sigma (O-C)^{2}$ (in {\it R})   &0.036 &0.055 &0.047 &0.132 &0.120 \\
\hline
\end{tabular}}
\end{center}
\end{table}

\begin{figure}
\centering
\includegraphics[trim = 0mm 0mm 0mm 0mm, clip, width=9cm]{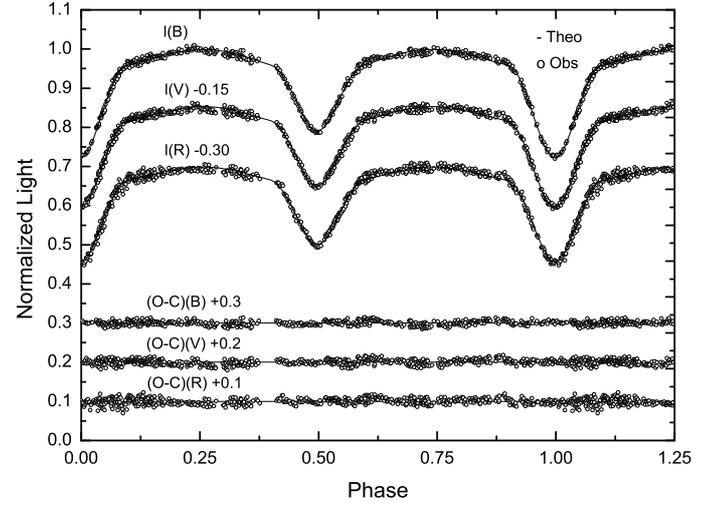}

\caption{The {\it BVR} light curves of ET Boo with the theoretical light curve solutions (solid lines). The final O-C residuals from the fit are also shown at the bottom of the figure.}
\label{fig1}
\end{figure}

\begin{figure}
\centering

\includegraphics[trim = 0mm 0mm 0mm 0mm, clip, width=9cm]{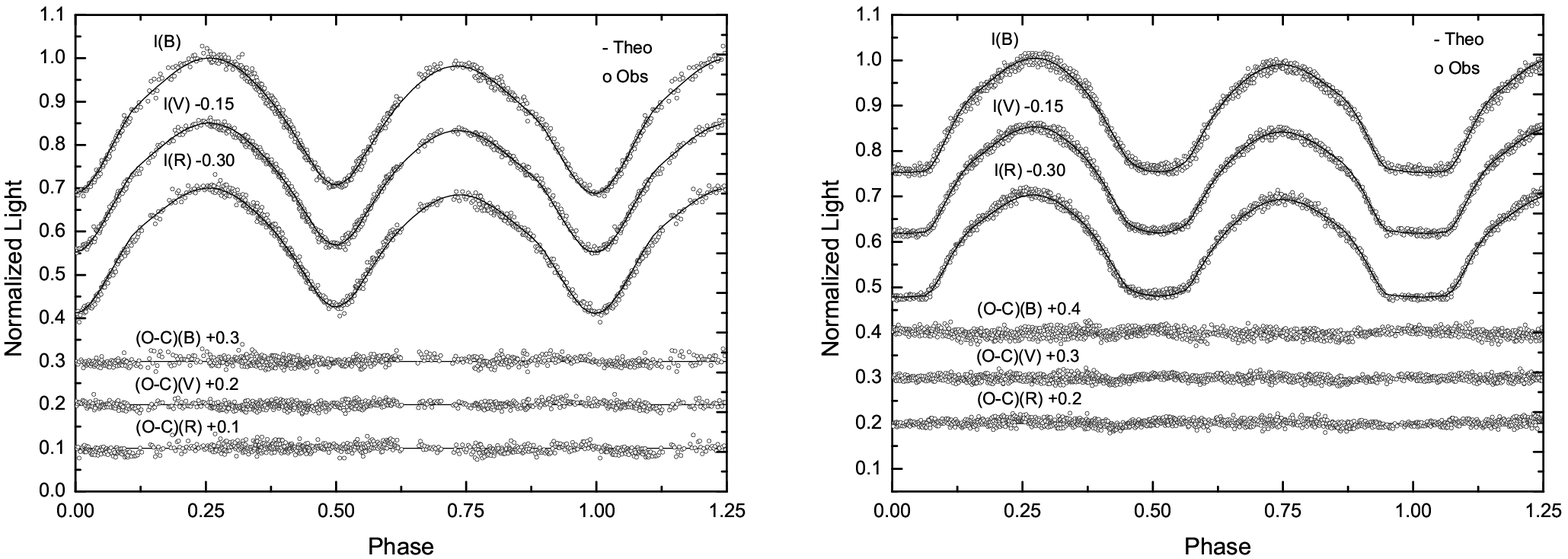}

\caption{The {\it BVR} light curves of V1123 Tau (left panel) and V1191 Cyg (right panel) with the theoretical light curve solutions (solid lines). The final O-C residuals from the theoretical fits are also shown at the bottom of each panel.}
\label{fig2}
\end{figure}

\begin{figure}
\centering

\includegraphics[trim = 0mm 0mm 0mm 0mm, clip, width=9cm]{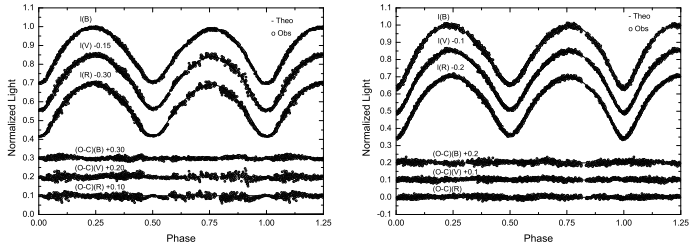}

\caption{The {\it BVR} light curves of V1073 Cyg (left panel) and V357 Peg (right panel) with the theoretical light curve solutions (solid lines). The final O-C residuals from the theoretical fits are also shown at the bottom of each panel.}
\label{fig3}
\end{figure}

The absolute parameters of five binary stars, obtained by means of the WD analyses, are given in Table~\ref{table5}. The optimum fit to each passband observed light curves (Obs) to the synthetic ones (Theo) are shown in Fig.~\ref{fig1} for \astrobj{ET Boo}, in Fig.~\ref{fig2} for \astrobj{V1123 Tau} and \astrobj{V1191 Cyg}, and in Fig.~\ref{fig3} for \astrobj{V1073 Cyg} and \astrobj{V357 Peg}. The final (O-C) residuals between the observed (Obs) and optimum synthetic light curves are also given in these three Figures.

\begin{table}
\scriptsize
\begin{center}
\caption{Absolute parameters, obtained by means of the WD analyses, of the five systems given in Table 4.}
\label{table5}
\medskip
\resizebox{9cm}{!} {\begin{tabular}{lccccc}
\hline
System & ET Boo & V1123 Tau & V1191 Cyg & V1073 Cyg & V357 Peg \\
\hline
a ($R_{\odot}$)         & 4.06$\pm$0.01 & 2.68$\pm$0.02 & 2.18$\pm$0.01& 4.70$\pm$0.02& 3.09$\pm$0.03 \\
$M_{h} (M_{\odot})$	& 1.15$\pm$0.02 & 0.35$\pm$0.01 & 0.14$\pm$0.01 & 1.73$\pm$0.10 & 0.85$\pm$0.03 \\
$M_{c} (M_{\odot})$	& 1.02$\pm$0.03 & 1.27$\pm$0.03 & 1.28$\pm$0.02 & 0.53$\pm$0.06 & 0.34$\pm$0.02 \\
$R_{h} (R_{\odot})$ 	& 1.096$\pm$0.003 & 0.78$\pm$0.03 & 0.48$\pm$0.05 & 2.33$\pm$0.11 & 1.48$\pm$0.13 \\
$R_{c} (R_{\odot})$	& 1.369$\pm$0.004 & 1.36$\pm$0.03 & 1.27$\pm$0.06 & 1.36$\pm$0.12 & 0.99$\pm$0.14 \\
$L_{h} (L_{\odot})$     & 1.51$\pm$0.01 & 0.66$\pm$0.04 & 0.32$\pm$0.07 & 9.77$\pm$0.95 & 4.73$\pm$0.85 \\
$L_{c} (L_{\odot})$	& 1.85$\pm$0.02 & 1.91$\pm$0.11 & 2.16$\pm$0.27 & 3.01$\pm$0.67 & 1.77$\pm$1.54\\
$Log g_{h} (cgs)$	& 4.42 & 4.20 & 4.22 & 3.94 & 4.02 \\	
$Log g_{c} (cgs)$	& 4.17 & 4.27 & 4.34 & 3.89 & 3.97 \\
$M_{bol, hot}$		& 4.30 & 5.19 & 5.98 & 2.28 & 3.06 \\
$M_{bol, cool}$		& 4.08 & 4.05 & 3.91 & 3.55 & 4.13 \\
\hline
\end{tabular}}
\end{center}
\end{table}

\section{Results and conclusions}

New CCD {\it BVR} observations of five eclipsing binaries (\astrobj{ET Boo}, \astrobj{V1123 Tau}, \astrobj{V1073 Cyg}, \astrobj{V1191 Cyg} and \astrobj{V357 Peg}) were obtained and the light curve analyses performed to acquire the absolute parameters of the components of these five systems. According to the results of simultaneous WD analysis given in Table~\ref{table4} and the absolute parameters given in Table~\ref{table5}, following evaluations can be inferred:

{\bf \astrobj{ET Boo}:}
Up to now, the absolute parameters of \astrobj{ET Boo} seem to be given for the first time in this study. The results given in Tables~\ref{table4} and \ref{table5} are obtained with the detached binary mode in WD code and no spots located on the components of \astrobj{ET Boo} binary system.

{\bf \astrobj{V1123 Tau}:}
The results given in Tables~\ref{table4} and \ref{table5} are obtained with the MODE 3 for overcontact binaries and one dark spot located on the cooler component of \astrobj{V1123 Tau}. The absolute parameters of \astrobj{V1123 Tau}, with their error values, obtained in this study (see Table~\ref{table5}) are consistent with those given by \citet{zahng11}, but they are a bit lower than those given by \citet{deb11} (see Table~\ref{table6}) due mainly to the different values of orbital inclination and third light contribution. \citet{deb11} obtained the orbital inclination and the third light level as \ i($^{\circ}$)= 68.10$\pm$0.24 \ and $L_{3} =0.013\pm0.009$ in {\it V} while our results gave the value of this parameters as $74.01\pm1.05$ and $0.186\pm0.002$, respectively. Also, \citet{deb11} have not attempted to obtain any spot parameters in their analysis.   

\begin{table}
\scriptsize
\begin{center}
\caption{Absolute parameters, of  \astrobj{V1123 Tau},  \astrobj{V1191 Cyg} and  \astrobj{V357 Peg} systems as found in the literature.}
\label{table6}
\medskip
\resizebox{9cm}{!} {\begin{tabular}{lccccc}
\hline
System & \multicolumn{2}{c}{V1123 Tau} & \multicolumn{2}{c}{V1191 Cyg} & V357 Peg \\
\hline
a ($R_{\odot}$)& - & 2.779$\pm$0.011&2.20$\pm$0.08& 2.194$\pm$0.012&3.920$\pm$0.016\\
$M_{h} (M_{\odot})$&0.40$\pm$0.02&0.392$\pm$0.015&0.13$\pm$0.01&0.139$\pm$0.08&0.690$\pm$0.013\\
$M_{c} (M_{\odot})$&1.36$\pm$0.05&1.404$\pm$0.070&1.29$\pm$0.08&1.306$\pm$0.022&1.720$\pm$0.015\\
$R_{h} (R_{\odot})$&0.80$\pm$0.01&0.789$\pm$0.004&0.52$\pm$0.15&0.518$\pm$0.003&1.250$\pm$0.024\\
$R_{c} (R_{\odot})$&1.37$\pm$0.02&1.389$\pm$0.006&1.31$\pm$0.18&1.307$\pm$0.007&2.120$\pm$0.015\\
$L_{h} (L_{\odot})$&0.67$\pm$0.04&0.695$\pm$0.129&0.46$\pm$0.25&0.463$\pm$0.006&2.406$\pm$0.333\\
$L_{c} (L_{\odot})$&2.01$\pm$0.07&2.130$\pm$0.385&2.71$\pm$0.80&2.731$\pm$0.029&9.673$\pm$1.195\\
References&(1)&(2)&(3)&(4)& (2)\\
\hline
\multicolumn{6}{c}{(1)\citet{zahng11}, (2)\citet{deb11}, (3)\citet{ulas12}, (4)\citet{zhu11}}\\
\end{tabular}}
\end{center}
\end{table}

{\bf \astrobj{V1191 Cyg}:}
The results given in Tables~\ref{table4} and \ref{table5} are obtained with the MODE 3 for overcontact binaries and one dark spot located on the cooler component of \astrobj{V1191 Cyg}. The absolute parameters of \astrobj{V1191 Cyg}, with their error values, obtained in this study (see Table~\ref{table5}) are consistent with those given by \citet{ulas12} and \citet{zhu11}(see Table~\ref{table6}). 

{\bf \astrobj{V1073 Cyg}:}
The results given in Tables~\ref{table4} and \ref{table5} are obtained with the MODE 3 for overcontact binaries and one dark spot located on the cooler component of \astrobj{V1073 Cyg}.  The results given by \citet{ahn92} on the masses, radii, temperatures, filling factor for \astrobj{V1073 Cyg} binary system were somewhat different from the results of this study. They found that the masses, radii, temperatures, and filling factor as $M_{h}=0.51\pm0.01 M_{\odot}$, $M_{c}=1.60\pm0.02 M_{\odot}$, $R_{h}=1.33\pm0.02 R_{\odot}$, $R_{c}=2.24\pm0.02 R_{\odot}$, $L_{h}=0.46\pm0.03 L_{\odot}$, $L_{c}=0.95\pm0.03 L_{\odot}$, and f=0.92, respectively. \citet{pribul03} gave f = 0.04 and $T_{c}=6650 K$ for \astrobj{V1073 Cyg} in their "Catalogue of the field contact binary stars". \cite{jafari06} also found $M_{h}=0.55 M_{\odot}$, $M_{c}=1.64 M_{\odot}$, $R_{h}=1.40 R_{\odot}$, $R_{c}=2.28 R_{\odot}$, $T_{h}=6700 K$, $T_{c}=6494 K$, f=0.19 with an unspotted model. In this study, the masses, radii, luminosities and the temperatures of the hotter and cooler components of \astrobj{V1073 Cyg} are obtained as listed in Table~\ref{table5}. 

\citet{dumitdin76} gave a light curve of the system obtained with no filter. But they did not present any kind of solution to this light curve. \citet{ahn92} obtained new Reticon spectral observations of the system and computed the mass ratio as 0.32$\pm$0.01. Because they determined a late spectral type for the system as noted by \citet{hildhil75} from the {\it uvby} colors and \citet{hill75}'s spectral classification (F2~IV-F1~V), they assumed convective envelopes for both of the components. \citet{sezer93} obtained photoelectric light curves in {\it B} and {\it V} bands, and used the WD computer program to find 3\% overcontact and a mass ratio of 0.436, assuming radiative behavior and $T_{h} = 8570 K$. \citet{sezer96} revised the analyses of his previous light curves, this time assuming convective behavior, and $T_{h} = 6700 K$, consistent with the spectroscopic values of \citet{ahn92} instead of taking A3Vm as the spectral type, and found 19\% - 22\% overcontact, a mass ratio of 0.306 and {\it i} = 69$^{\circ}$.4. The light curves of the system were reported also to be variable by \citet{yanli00}. While a positive O'Connell effect had been observed in 1963-1964 \citep{kondo66}, a negative O'Connell effect was observed in 1988-1991 \citep{sezer93}. \citet{yanli00} could not find a definite solution for the unstable behavior of the light curves of \astrobj{V1073 Cyg} and they grouped it with AU Ser and FG Hya as they are A-subtype W UMa stars showing instability in both their orbital periods and light curves.

{\bf \astrobj{V357 Peg}:}
The results given in Tables~\ref{table4} and \ref{table5} are obtained with the MODE 3 of WD code for overcontact binaries and one bright spot located on the hotter component of \astrobj{V357 Peg} binary system. The absolute parameters of \astrobj{V357 Peg}, with their error values, obtained in this study (see Table~\ref{table5}) were somewhat different from those given by \citet{deb11}(see Table~\ref{table6}). The main difference between the WD light curve analysis by \citet{deb11} and the analaysis of this study is the bright spot we located on the hotter component of \astrobj{V357 Peg} system(see Table~\ref{table4}). Therefore, this hot spot may indicate the effect of a mass transfer between the components of the system and this effect may be the cause of somewhat different values of absolute parameters that we have in our analysis.  

{\bf MK classifications and evolutionary states:}
An evaluation of the log~g and $log T_{ph}$ values of the components of five eclipsing binaries(\astrobj{ET Boo}, \astrobj{V1123 Tau}, \astrobj{V1073 Cyg}, \astrobj{V1191 Cyg} and \astrobj{V357 Peg}) in the $log~g - log T_{e}$ diagrams given by \citet{strai81} and by \citet{maemey88} reveals the following results: 

The spectral types of both components of \astrobj{ET Boo} could be F7-8~IV or MK classification for \astrobj{ET Boo} could be as F8~V+F8~V which is not far from one another(aside from luminosity classes IV). In a sense, this estimation was a verification of the F8 spectral type of the system given by \citet{bart02}. No theoretical study on the evolutionary status on detached \astrobj{$\beta$ Lyr} type close binary systems has been published yet. Therefore, we could not evaluate the absolute parameters of the components of \astrobj{ET Boo} to see and examine the evolutionary characteristics of the component stars of \astrobj{ET Boo}.
However, on the occasion of the similarity of \astrobj{ET Boo} and some short period \astrobj{RS CVn} type binaries (e.g. \astrobj{CG Cyg}, \astrobj{WY Cnc}, \astrobj{RT And}, and \astrobj{ER Vul}), in point of their mass ratio, and absolute parameters \citep[see][]{dryo05,bud96,last02,kjur03,kjur04}, a plausible inference can be made for the relationship of the evolutionary status of \astrobj{ET Boo} system. \citet{last02} showed that \astrobj{RT And} and \astrobj{CG Cyg} have a secondary component far too cool to be matched by the same isochrone as the primary. They also pointed out to the same difficulty with the models by fitting simultaneously their effective temperatures, masses and radii \citep[see][]{pols97}. In addition to effective temperature revision, they gave a possible explanation of the disagreement that may come from mass transfer and starspot activity. Therefore, it can be inferred that \astrobj{ET Boo} system is more likely to undergo an evolutionary progress to be a short period RS CVn-type system than to be a W UMa-type contact binary. In order to confirm this point of view, it will be better to take some high resolution spectra of \astrobj{ET Boo} to evaluate this point together with the activity phenomena.

The spectral types of the components of \astrobj{V1123 Tau} could be G0~V+G1-2~V or MK classification for the cooler component of \astrobj{V1123 Tau} could be as G1-2~V. However, \citet{rucin08} estimated that G0~V is the spectral type of \astrobj{V1123 Tau}.

The spectral types of the components of \astrobj{V1191 Cyg} could be F6-7~V-IV or MK classification for the cooler component of \astrobj{V1191 Tau} could be $\sim$ F6-7V. This prediction is in agreement with the estimation by \citet{rucin08} on the spectral type of \astrobj{V1191 Cyg} as F6~V.

The spectral types of the components of \astrobj{V1073 Cyg} could be F5~IV + F5~IV or MK classification for the hotter component of \astrobj{V1073 Tau} could be $\sim$ F5 V - IV. But the results given by \citet{ahn92} on MK spectral type for \astrobj{V1073 Cyg} binary system were somewhat different from the results of this study. They found the spectral type as F1-F2. \citet{hill75} gave also the spectral classification as F2~IV-F1~V.  

The spectral types of the components of \astrobj{357 Peg} could be F2~IV+F3~IV. However, \citet{rucin08} estimated that F2V is the spectral type of \astrobj{V357 Peg}. 

\citet{hild88} showed that some primary components of the contact W-type W UMa systems are located below the ZAMS in the mass-luminosity diagram due to luminosity transfer to the secondary components, and the secondary components of the W-type systems all have larger radii than expected for their ZAMS masses. On the other hand, the A-type W UMa systems which are more evolved than W-type counterparts, are located near or beyond the TAMS. And the secondary components of these A-type systems have substantially larger radii than expected for their ZAMS masses. By comparing and evaluating the masses, radii, luminosities and the temperatures of W-type systems contact binaries (\astrobj{V1123 Tau}, \astrobj{V1191 Cyg}), it can be seen that the characteristics of hotter and cooler components of \astrobj{V1123 Tau} and \astrobj{V1191 Cyg} are in agreement with the results of \citet{hild88} on the secondary and primary components of W-type systems, respectively. Again, with the values of masses, radii, luminosities and temperatures of the hotter and cooler components of A-type system contact binaries (\astrobj{V1073 Cyg}, \astrobj{V357 Peg}) it can also seen that the characteristics of hotter and cooler components of \astrobj{V1073 Cyg} and \astrobj{V357 Peg} are in agreement with the results of \citet{hild88} on the primary and secondary components of A-type systems, as well. 

Consequently, the masses and the radii of the components together with their bolometric absolute magnitudes, given in Table~\ref{table5}, are important indicators for the evolutionary status of the components in the context of close binary evolution.

\section{Acknowledgments}
This research has made use of the Simbad database, operated at CDS, Strasbourg, France, and of NASA's Astrophysics Data System Bibliographic Services. \"{O}B would like to thank The Scientific and Technological Research Council of Turkey (T\"{U}B\.{I}TAK) for their support through BIDEB-2211 scholarship.

\end{document}